\documentclass[prl,superscriptaddress,showpacs,twocolumn,longbibliography]{revtex4-1}

\usepackage{hyperref}

\usepackage{color}
\usepackage[usenames,dvipsnames]{xcolor}
\usepackage{amsmath,amsthm,amssymb}
\usepackage{graphicx}
\usepackage{epsfig}
\usepackage{dcolumn}
\usepackage{bm}
\usepackage{mathrsfs}
\usepackage{multirow}
\usepackage[all]{xy}
\usepackage{pbox}
\usepackage{verbatim}
\usepackage{bbold}
\usepackage{braket}

\def\(({\left(}
\def\)){\right)}
\def\[[{\left[}
\def\]]{\right]}

\newcommand{\be}{\begin{equation}}
\newcommand{\ee}{\end{equation}}
\newcommand{\ben}{\begin{eqnarray}}
\newcommand{\een}{\end{eqnarray}}
\newcommand{\beq}{\begin{equation}}
\newcommand{\eeq}{\end{equation}}

\begin{document}

\title{Quantum contact process}

\author{Federico Carollo}
\affiliation{School of Physics and Astronomy, University of Nottingham, Nottingham, NG7 2RD, UK}
\affiliation{Centre for the Mathematics and Theoretical Physics of Quantum Non-Equilibrium Systems,
University of Nottingham, Nottingham, NG7 2RD, UK}

\author{Edward Gillman}
\affiliation{School of Physics and Astronomy, University of Nottingham, Nottingham, NG7 2RD, UK}
\affiliation{Centre for the Mathematics and Theoretical Physics of Quantum Non-Equilibrium Systems,
University of Nottingham, Nottingham, NG7 2RD, UK}

\author{Hendrik Weimer}
\affiliation{Institut f\"ur Theoretische Physik, Leibniz Universit\"at Hannover, Appelstra\ss e 2, 30167 Hannover, Germany}

\author{Igor Lesanovsky}
\affiliation{School of Physics and Astronomy, University of Nottingham, Nottingham, NG7 2RD, UK}
\affiliation{Centre for the Mathematics and Theoretical Physics of Quantum Non-Equilibrium Systems,
University of Nottingham, Nottingham, NG7 2RD, UK}

\date{\today}

\begin{abstract}
The contact process is a paradigmatic classical stochastic system displaying critical behavior even in one dimension. It features a non-equilibrium phase transition into an absorbing state that has been widely investigated and shown to belong to the directed percolation universality class. When the same process is considered in a quantum setting much less is known. So far mainly semi-classical studies have been conducted and the nature of the transition in low dimensions is still a matter of debate. Also from a numerical point of view, from which the system may look fairly simple --- especially in one dimension --- results are lacking. In particular the presence of the absorbing state poses a substantial challenge which appears to affect the reliability of algorithms targeting directly the steady-state. Here we perform real-time numerical simulations of the open dynamics of the quantum contact process and shed light on the existence and on the nature of an absorbing state phase transition in one dimension. We find evidence for the transition being continuous and provide first estimates for the critical exponents. Beyond the conceptual interest, the simplicity of the quantum contact process makes it an ideal benchmark problem for scrutinizing numerical methods for open quantum non-equilibrium systems.
\end{abstract}

\maketitle

\noindent {\bf \em Introduction -- } Understanding the non-equilibrium behavior of many-body quantum systems is one of the major goals of current research in physics. From the experimental side, recent technological developments and increased capabilities in the realisation and control of quantum systems offer promising platforms for the investigation of quantum phenomena far from equilibrium \cite{Syassen1329,Kim2010,Barreiro2011,MULLER2012,Schreiber2015,Eisert2015,Islam2015,Bohnet2016,Luschen2016,lienhard2017,Lebrat2018,Wade2018}. However, from a theoretical perspective, non-equilibrium quantum systems are typically much more complex than classical ones and their characterization is still an open problem, especially when going beyond the realm of exactly solvable models or the application of semi-classical approaches \cite{Prosen2008,Prosen2008b,Znidaric2010,Clark2010,Prosen2011,Prosen2011c,Karevski2013,Ilievski2017}. Even numerical studies, which in the classical case allow for the accurate investigation of non-integrable systems, in quantum settings are often severely limited due to computational constraints.

\begin{figure}[t]
\centering
\includegraphics[scale=0.24]{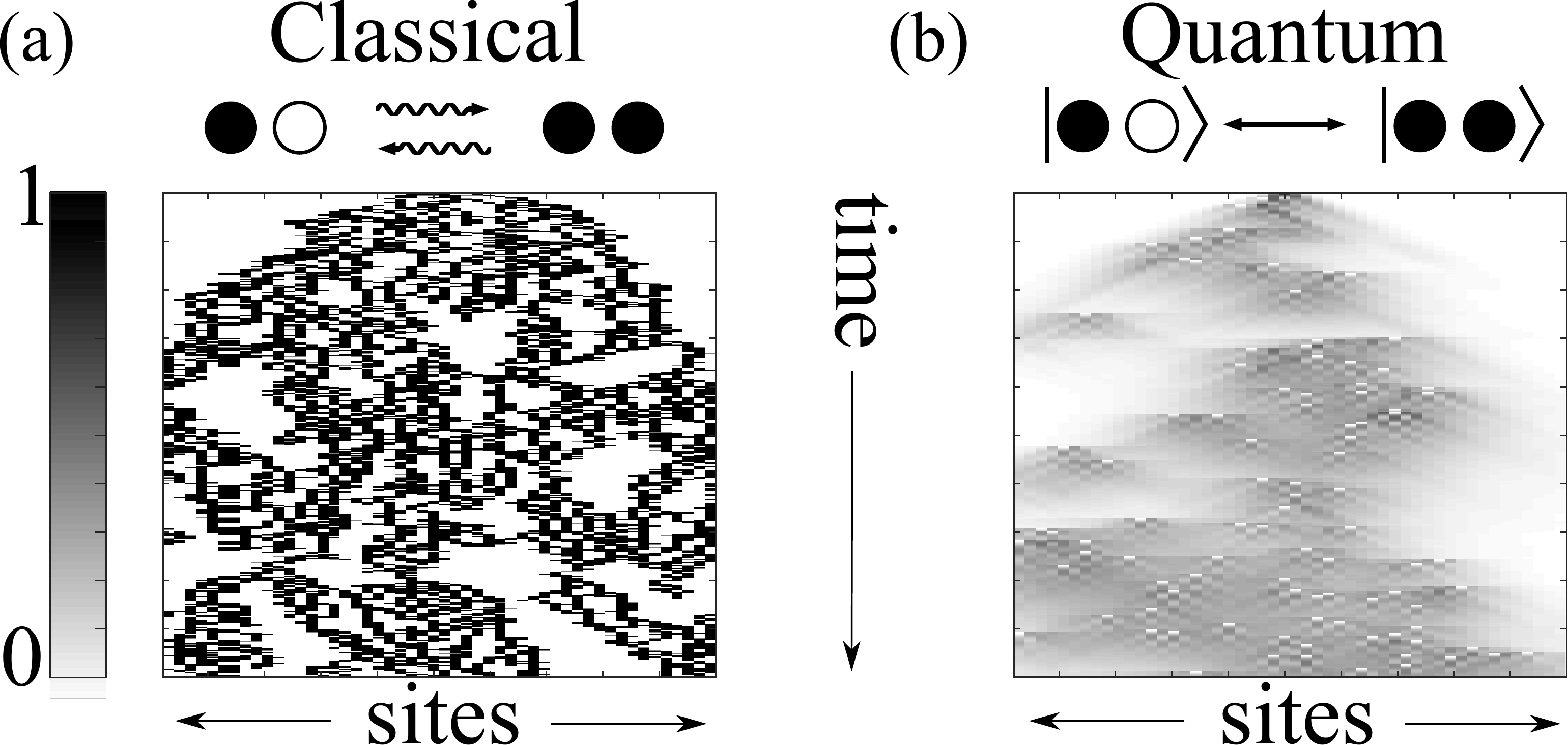}
\caption{\textbf{Classical vs. quantum dynamics.} (a) Trajectory with incoherent branching (coagulation) process: an empty (occupied) site can become occupied (empty) if at least one of the neighboring sites is occupied. Within a given trajectory sites are either empty or occupied. (b) Trajectory of the quantum process, where branching (coagulation) is driven coherently by a Hamiltonian. Superposition of different configurations are generated within a single trajectory and the density hence assumes values that are in between zero and one.}
\label{FigR1}
\end{figure}

A paradigmatic example --- illustrating the gap in our understanding of classical and quantum non-equilibrium systems --- is the {\it contact process}. A contact process model deals with a $d$-dimensional lattice system, whose sites can be either empty or occupied by a particle. In a classical setting, the dynamics is given by two incoherent processes: (i) {\it self-destruction}, consisting of spontaneous particle decay, and (ii) {\it branching} ({\it coagulation}) for which an empty (occupied) site can become occupied (empty) only if at least one particle is present in the neighboring sites [{\it cf.} Fig.~\ref{FigR1}(a)]. The non-equilibrium behavior resulting from these dynamical rules has been widely investigated, as it is relevant for e.g. epidemic spreading or growth of bacterial colonies \cite{mollison1977,Grassberger1983,Kuhr2011}, and these systems exhibit second-order absorbing state phase transitions, which belong to the {\it directed percolation} (DP) universality class \cite{dickman1991,Jensen1993,Hinrichsen2000b}.

When {\it branching/coagulation} is treated as a quantum process, see Fig.~\ref{FigR1}(b), non-classical effects, such as superpositions, entanglement and quantum correlations, come into play. A comprehensive understanding of the non-equilibrium physics in this setting, potentially realizable in cold atom experiments \cite{Gutierrez2017,Buchhold2017,Helmrich2018,Helmrich2018b}, still remains elusive. Using mean-field theory approximations, i.e. neglecting correlations among sites, it was found that the absorbing state phase transition of the contact process survives in the quantum regime \cite{Marcuzzi2016,Buchhold2017,MinjaeJo2019}. However, these results also suggest that the transition is of first-order for any lattice dimension. This contradicts the common belief that discontinuous transitions should not occur in generic non-equilibrium $1d$ systems with fluctuating ordered phases \cite{Hinrichsen2000} and a softening of the transition into a second-order one has been conjectured \cite{Roscher2018}. This clearly shows that efforts to capture the physics of the quantum contact process must go beyond semi-classical approaches and use techniques which leave quantum correlations between different sites intact.

In this paper, we conduct a detailed investigation of the $1d$ quantum contact process (QCP) and make substantial steps towards a comprehensive understanding of the critical behavior of this challenging quantum non-equilibrium problem. Applying the {\it infinite time-evolving block decimation} (iTEBD) algorithm, we find evidence for a continuous non-equilibrium phase transition. We also establish an estimate for the location of the critical point and provide estimates for the critical exponents, which suggest that the universality class is in fact different from that of (classical) directed percolation.

\noindent {\bf \em The model -}
To model the QCP for a finite size system, we consider a chain of $L$ sites with open boundaries. Attached to each site there is a two-level system with basis $\{|\bullet\rangle , |\circ\rangle  \}$, representing occupied and empty sites respectively. The evolution of the quantum state $\rho(t)$ is governed by Lindblad generators \cite{Gorini1976,Lindblad1976,Breuer2002,Gardiner2004}, $\dot{\rho}(t)=\mathcal{L}[\rho(t)]$, with $\mathcal{L}[\rho]=-i[H,\rho]+\mathcal{D}[\rho]$. The purely dissipative contribution to the dynamics, $\mathcal{D}[\rho]$, encodes the (classical) spontaneous particle decay ($|\bullet \rangle  \to |\circ\rangle $):
\begin{align}
\mathcal{D}[\rho]=\gamma\sum_{k=1}^L\left(\sigma_-^{(k)}\rho\sigma_+^{(k)}-\frac{1}{2}\left\{n^{(k)},\rho\right\}\right)\, ,
\end{align}
where $\sigma_-|\bullet \rangle =|\circ\rangle$, $\sigma_-|\circ\rangle=0$, and $\sigma_+=\sigma_-^\dagger$ are the ladder spin operators, while $n=\sigma_+\sigma_-$ is the number operator. The Hamiltonian $H$, instead, encodes the coherent version of the branching/coagulation process and has the form \cite{Marcuzzi2016,Buchhold2017,Roscher2018} (with $\sigma_1|\bullet/\circ\rangle=|\circ/\bullet\rangle$),
\begin{align}
H=\Omega \sum_{k=1}^{L-1}\left( \sigma_1^{(k)}n^{(k+1)}+ n^{(k)}\sigma_1^{(k+1)}\right)\, .
\label{H}
\end{align}
The structure of this Hamiltonian is such that the state of a site can (coherently) evolve only if at least one of the neighboring sites is occupied.

By construction, the vacuum state $\rho_{\rm 0}=\ket{0} \hspace{-0.1cm}\bra{0}$, with $ \ket{0}=\bigotimes_{k=1}^L|\circ\rangle$, is a steady-state of the open system dynamics. This is an absorbing state, characterized by zero dynamical fluctuations. In the classical contact process, an additional steady-state can emerge in the thermodynamic limit and for sufficiently large values of the branching rate. This state has a finite density of particles, and at a critical branching rate one observes a non-equilibrium absorbing state phase transition, which belongs to the DP universality class \cite{dickman1991,Jensen1993,Hinrichsen2000b}.

\noindent {\bf \em Numerical methods -} The numerical simulation of quantum dissipative dynamics remains a major challenge \cite{Bonnes2014,Weimer2015,Florian2016,Jin2016,Werner2016,Chen2018,Jaschke2018b,Raghunandan2018}. Exact diagonalization of the Lindblad generator is limited to small systems for which the transition in the QCP cannot be detected.
To study larger systems \cite{Prosen2009,Ljubotina2017}, one needs to resort to approximate representations of the quantum state, e.g. through {\it matrix product states} (MPSs) \cite{SCHOLLWOCK2001,Paeckel2019}. A possible way to study a non-equilibrium phase transition is thus offered by MPS techniques targeting the steady-state of the dynamics \cite{Cui2015,Gangat2017}. However, in the case of non-equilibrium phase transitions, universal information is also contained in the dynamics itself. Furthmore, for our model, we observe that these methods struggle to pinpoint the transition, as they tend to be biased towards the uncorrelated (absorbing) steady-state, $\rho_{\rm 0}$, in any parameter regime. For these reasons, we run real-time dynamical simulations by means of TEBD algorithms \cite{Vidal2004,Zwolak2004,Vidal2007,Pfeifer2014}, working in the thermodynamic limit (iTEBD). These simulations directly implement the open system dynamics in Liouville space \cite{Orus2008,Kshetrimayum2017}. Their accuracy is limited by a finite bond dimension $\chi$ in the approximation of the evolved state. Nonetheless, this strategy currently seems to be the only one possible, among existing algorithms, to study the QCP. For our simulations we used different Trotter schemes with time-steps (time is given in units of $\gamma^{-1}$, throughout), $0.01\le dt\le 0.1$, and $\chi\le 1300$, reaching simulation times $t\approx 50$ on standard PCs. 

\begin{figure}[t]
\centering
\includegraphics[scale=0.275]{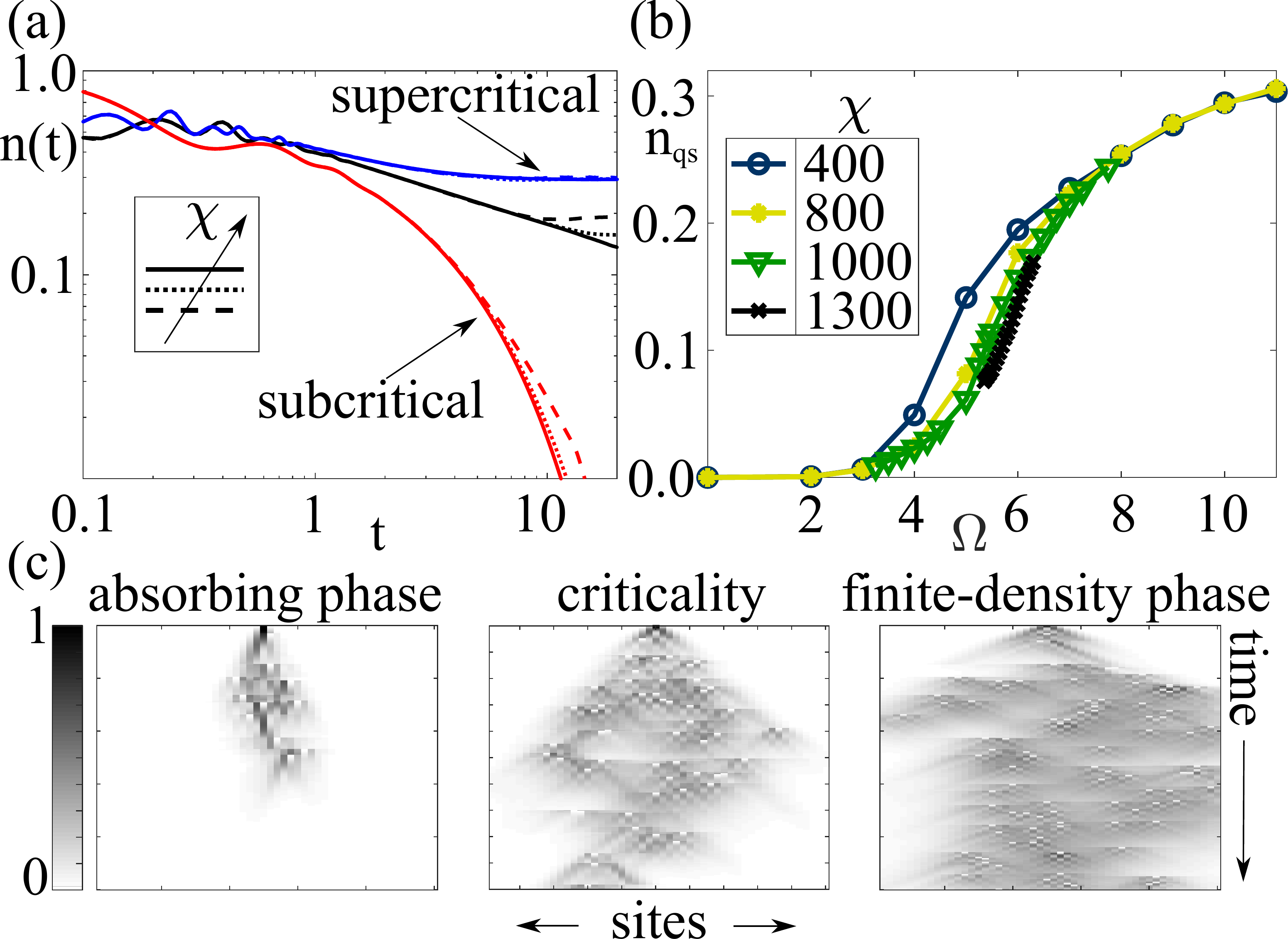}
\caption{\textbf{iTEBD simulations.} (a) Log-log plot of the average density $n{(t)}$. For $\Omega=2$ (bottom curves with $\chi=50,100,200$), a rapid convergence to an exponentially decaying curve is observed. For $\Omega=10$ (top curves with $\chi=200,400,800$) instead a convergence to a stationary finite density curve is shown. Curves in the middle are for $\Omega=6$: here, larger bond dimensions ($\chi=350,1000,1300$) are needed because of entanglement and correlation growth in the MPS close to criticality. (b) Quasi-stationary density $n_{\rm qs}$ (taken at $t=20$): its dependence on $\Omega$ seems to suggest a continuous transition. (c) Density-plot of the site average occupation for representative quantum trajectories starting from a single seed. The initial state is the one with a single occupied site. Results from TEBD simulations with $L=50$ and $\chi=300$. Times and rates are in units of $\gamma^{-1}$ and $\gamma$, respectively.}
\label{FigR2}
\end{figure}
\noindent {\bf \em Non-equilibrium phase transition - } As a first step, we establish the existence of an absorbing state phase transition in the QCP and estimate the location of the critical point, $\Omega_{c}$. The order parameter is the average number of particles,  $n(t)=L^{-1}\sum_k {\rm Tr}\left(\rho(t)n^{(k)}\right) $, which is zero in the absorbing phase, $\Omega < \Omega_{c}$, and non-zero in the finite-density phase $\Omega > \Omega_{c}$.

To establish these phases numerically we estimate the quasi-stationary density, $n_{\rm qs}$, by using the approximation of $n(t)$ obtained from the iTEBD simulations, for sufficiently large values of $t$ where the density is approximately stationary, see Fig. \ref{FigR2}(a). 
With a bond dimension of $\chi \le 800$ we can establish convergence of $n(t)$ up to $\Omega \approx 3$ [see Fig. \ref{FigR2}(b)]. For these values, we find that the density $n(t)$ decays exponentially to zero and $n_{\rm qs}\approx 0$. Thus, this region belongs to the absorbing phase.
For larger $\Omega$ it becomes challenging to establish convergence (see discussion further below). However, when $\Omega \ge 8$ it again becomes possible to approximate $n_{\rm qs}$ with $\chi \le 800$. The density $n_{\rm qs}$ is found to be non-zero, which establishes the existence of an active phase and pins down the phase transition region to the interval $3 < \Omega_{c} < 8$. Here long relaxation times together with the concomitant buildup of entanglement and quantum correlations require a large MPS bond dimension for establishing convergence. This is visible in the stationary density displayed in Fig.~\ref{FigR2}(b), whose shape is suggestive of a second-order phase transition, smoothed by finite time effects.

The different phases of the QCP become also visible in single dynamical realizations starting from a single seed, displayed in Fig. \ref{FigR2}(c). In the absorbing phase the seed creates a small cluster which does not spread and rapidly dies. In the active phase, instead, the branching process is dominant and the initial seed spreads populating the whole system. In the critical region, a contained propagation can be observed, and the density still tends to zero for later times. 

\noindent {\bf \em Critical behavior and exponents: universal dynamics -} We start the analysis of the critical region by studying dynamical observables. 
The location of the critical point $\Omega_{c}$ can be established by analyzing the time evolution of $n(t)$, as shown in Fig. \ref{FigR2}(a): the concavity of this curve, in a log-log plot, can indicate whether the corresponding $\Omega$ is supercritical (positive concavity) or subcritical (negative concavity). This allows us to improve the estimate for $\Omega_{c}$, particularly when combined with the observation that the finite bond-dimension effects consistently lead to an artificial saturation of $n(t)$ at a finite value [{\it cf.} Fig. \ref{FigR2}(a)-(b)]. Therefore, once a value of $\Omega$ is established to be in the absorbing phase for some $\chi$, we can be confident that it is also so as $\chi \to \infty$. This makes it possible establish a lower-bound on $\Omega_{c}$ by simply taking the highest $\chi$ simulations available and checking which $n(t)$-curves lie in the inactive phase. Of course, this also means that establishing an upper bound is more challenging and from analyzing our data we conclude that $\Omega_{c}$ must be in the range $\Omega_{c}\in [5.95,7]$.

\begin{figure}[t]
\centering
\includegraphics[scale=0.2680]{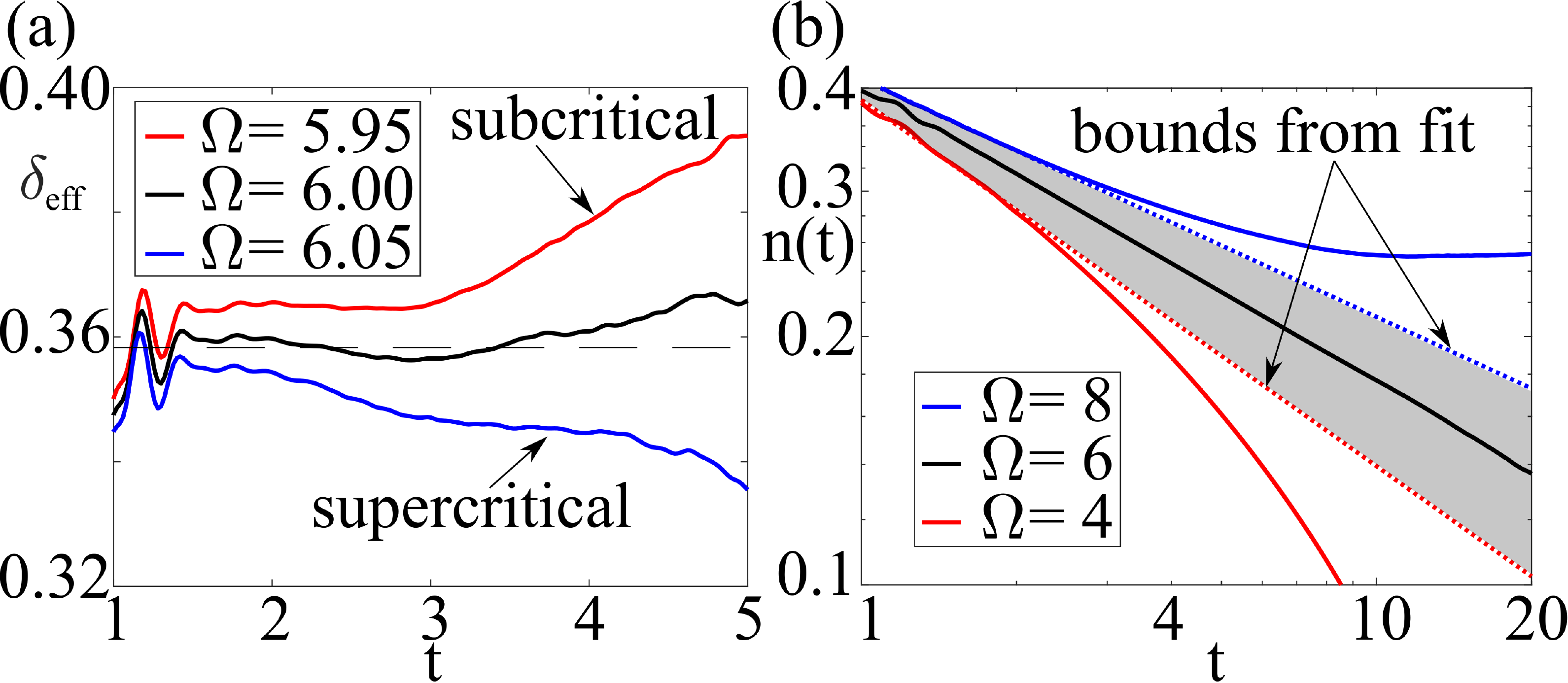}
\caption{\textbf{Determination of $\delta$-exponent.} (a) Plot of the effective exponent $\delta_{\rm eff}(t)$ of Eq.~\eqref{effexp}, with $b=4$. For  $\Omega=5.95$ the exponent increases at later times signaling that the curve is subcritical; this provides a lower bound for $\Omega_c$. On the contrary, $\Omega=6.05$ looks supercritical. For $\Omega=6$ we observe an almost constant behavior; we thus consider $\Omega_c\approx 6$. The dashed line guides the eye to the value $\delta=0.358$, obtained by averaging the latter curve for $t\in[1.5,3]$. (b) An uncertainty range for $\delta$ is obtained by extrapolating an algebraic behavior from two converged (in $\chi$) curves bounding $\Omega_c$. The fit is performed for $t\in[1,2]$, providing $\delta\in[0.28,0.44]$. Times and rates are in units of $\gamma^{-1}$ and $\gamma$, respectively.}
\label{FigR3}
\end{figure}

Under the assumption of a second-order absorbing state phase transition, the order parameter is expected to follow the universal scaling relation \cite{Hinrichsen2000b}
\begin{equation}
n(t) \approx t^{-\delta}\, f\left(\left(\Omega-\Omega_c\right) t^{1/\nu_\parallel}\right)\, ,
\label{UniDyn}
\end{equation}
where $\delta$ and $\nu_\parallel$ are critical exponents and $f$ a universal scaling function. The exponent $\nu_\parallel$ is related to the divergence of the time-correlations, while $\delta$ determines the critical algebraic decay of the density, $n(t) \approx t^{-\delta} f\left(0\right)$, for $\Omega=\Omega_c$. To obtain an estimate for the critical point $\Omega_c$ we search for the $n(t)$-curve showing algebraic decay [see critical regime in Fig. \ref{FigR2}(a)]. This is done by defining an effective exponent \cite{Hinrichsen2000b},
\begin{align}
\delta_{\rm eff}(t)=-\frac{1}{\log b}\log\frac{n(t\, b)}{n(t)} ~,
\label{effexp}
\end{align}
and identifying the $\Omega$-value for which $\delta_{\rm eff}(t)$ is as close as possible to a constant. In this way, as is shown in Fig.~\ref{FigR3}(a), we can recover our best estimate both for the critical rate $\Omega_{c}\approx 6$ and for the exponent $\delta \approx 0.36$. To provide bounds on the latter value, we extrapolate an algebraic behavior from a converged (in bond dimension) subcritical curve and a supercritical one: taking $\Omega=4$ to be the subcritical and $\Omega=8$ to be the supercritical one, as shown in Fig.~\ref{FigR3}(b), we obtain $\delta=0.36\pm 0.08$.

\begin{figure}[t]
\centering
\includegraphics[scale=0.34]{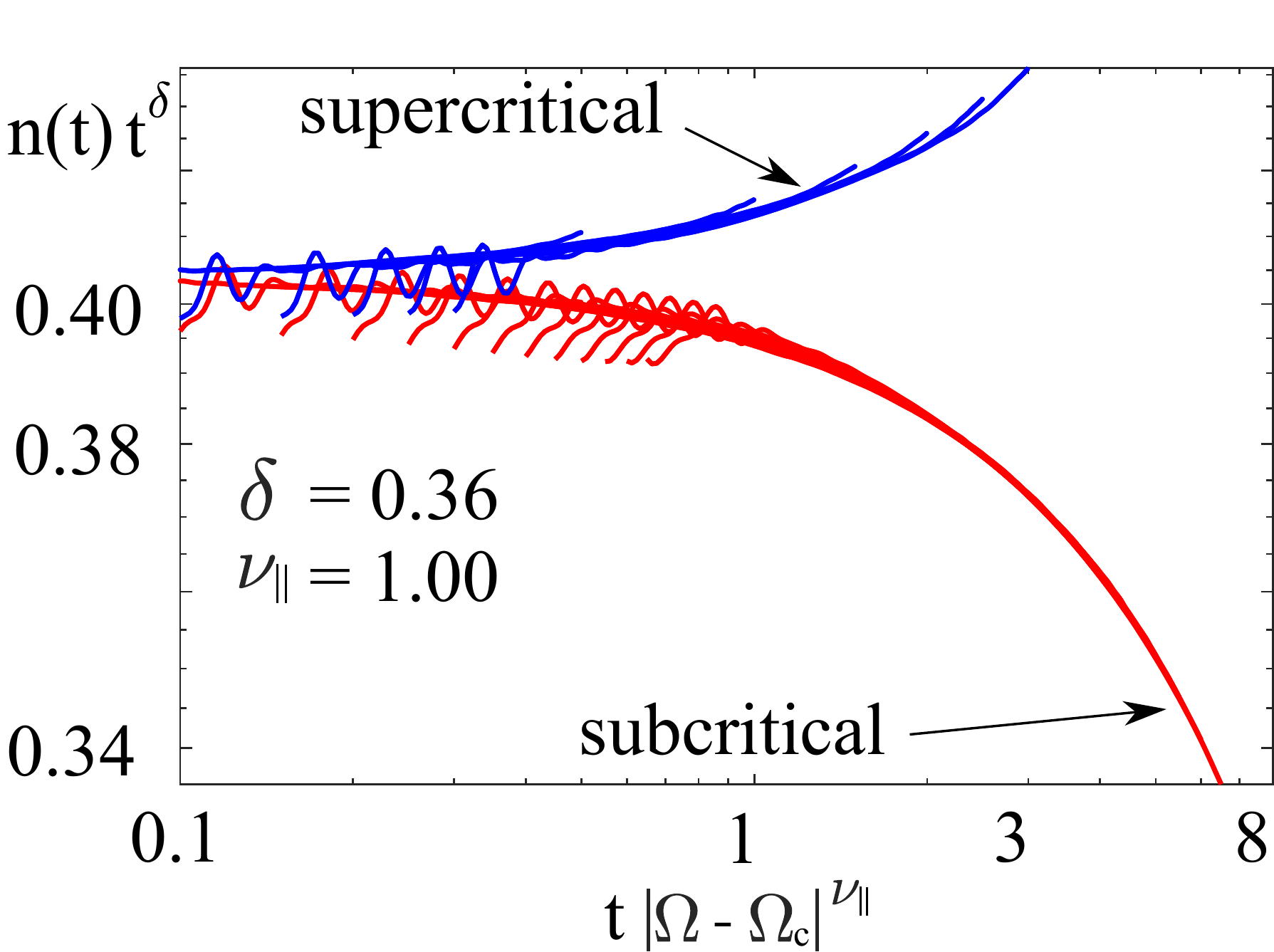}
\caption{\textbf{Determination of $\nu_\parallel$-exponent.} Scaled average density, $n(t)\, t^\delta$, as a function of $t\, |\Omega-\Omega_c|^{\nu_\parallel}$ for $\chi=1300$. Choosing the critical values $\delta=0.36$, $\nu_\parallel=1$ and $\Omega_c=6$ collapses all curves into two master curves. The detachment of the supercritical curves for long times from the master one is likely a finite bond dimension effect: low bond dimensions appear to cause an artificial density saturation. Times and rates are in units of $\gamma^{-1}$ and $\gamma$, respectively.}
\label{FigR5}
\end{figure}

The dynamical scaling relation \eqref{UniDyn} also implies that plotting $n(t)\, t^\delta$ as a function of $t\, |\Omega-\Omega_c|^{\nu_\parallel}$ should yield a collapse of all curves into two master curves, depending on whether the value of $\Omega$ is above or below critical. Fig.~\ref{FigR5} indeed shows such a collapse. This allows us to estimate $\nu_\parallel\approx 1$, and is a further hint towards the continuous nature of the phase transition.

\noindent {\bf \em Critical behavior and exponents: steady-state universality -} Estimates for static critical exponents are obtained by analyzing the density $n_{\rm qs}$ [Fig. \ref{FigR2}(b)] and the spatial correlation length $\xi_\perp$ near the critical point. These quantities are expected to behave as $n_{\rm qs}\approx |\Omega-\Omega_c|^\beta$ and $\xi_\perp\approx |\Omega-\Omega_c|^{-\nu_\perp}$, respectively, defining the critical exponents $\beta$ and $\nu_\perp$.

The determination of static exponents is more challenging than that of the dynamical ones, as it requires a large number of simulations up to long times. Additionally, while the exponent $\delta$ can be bound easily from a single active and inactive realization of $n(t)$ [Fig. \ref{FigR3}(b)], the values of $\beta$ and $\nu_\perp$ are extracted from fits that are highly sensitive to the considered region of $\Omega$-values. The determination of $\nu_{\perp}$ is particularly demanding as the correlation length constitutes a non-local observable, which is computed from the asymptotic behavior of the density-density correlation function, $C(r)=\braket{n^{(r)}n^{(0)}}-\braket{n^{(0)}}^2\sim e^{-r/\xi_{\perp}}$. In the vicinity of critical points in second-order phase transition where long-range correlations are expected, such non-local observables are difficult to approximate, since MPSs only support a maximal correlation length set by the bond dimension.

\begin{figure}[t!]
\centering
\includegraphics[scale=0.27]{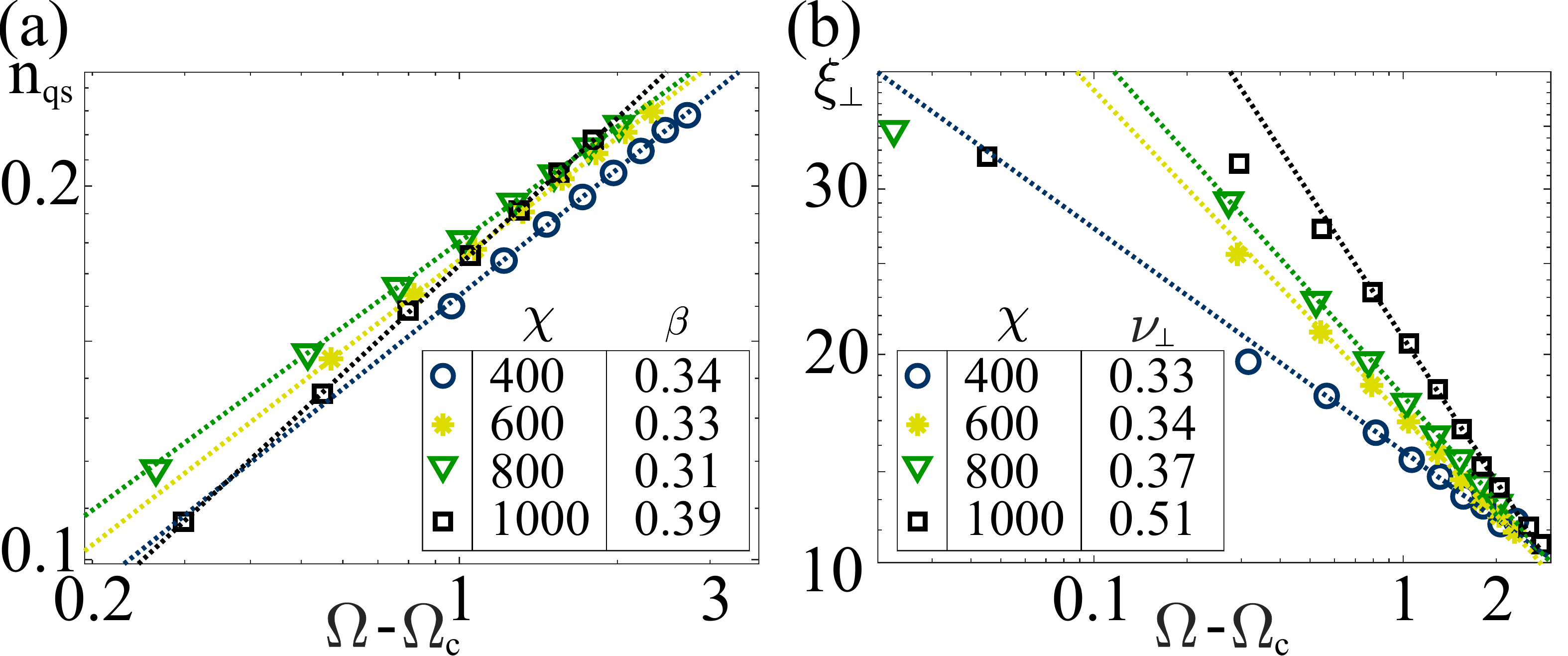}
\caption{\textbf{Steady-state exponents ($\beta$, $\nu_\perp$).} Plots in log-log scale with $\Omega$ in units of $\gamma$. (a) Extrapolation of $\beta$ from the quasi-stationary density (taken at $t=40$). Dashed lines are fits with power-law curves in the range $\Omega\in[5.5,7]$. We can estimate $\beta=0.39\pm0.08$: the value is obtained from the simulations with largest $\chi$, while the error is the maximal distance of this from lower bond dimensions. (b) Correlation length $\xi_\perp$: the behavior is consistent with second-order phase transition as correlations increase in the critical region. Fits are obtained by discarding the data showing a rounding off of the correlations length due to finite $\chi$ effects. }
\label{FigR4}
\end{figure}

To determine $\beta$  we investigate the range $\Omega\approx[5,7]$, performing a power-law fit with $\beta$ and $\Omega_{c}$ treated as free parameters, see Fig.~\ref{FigR4}(a). As $\chi$ is increased, the estimate of $\Omega_{c}$ is shifted to higher values, which is consistent with the artificial saturation of the density due to a finite bond dimension. Our estimate for the exponent is $\beta=0.39\pm 0.08$, with error given by the largest distance of the estimate from lower bond-dimension results. This is compatible with the value of $\beta$ that is obtained under the assumption that the hyperscaling relation \cite{Hinrichsen2000b} of the DP universality class also applies to the QCP: $\beta=\delta \nu_\parallel\approx 0.36$.

Fig.~\ref{FigR4}(b) displays the behavior of the correlation length $\xi_{\perp}$ as estimated from the density-density correlation function. All data are consistent with a power-law behavior. Near the critical point the correlation length systematically increases with increasing bond dimension. This suggests that, for a large enough $\chi$, one should be able to observe, close to criticality, diverging correlation lengths as expected in continuous transitions. However, our data are strongly affected by finite bond dimensions: near the critical point the correlation length rounds off instead of diverging. This is due to the fact that a finite bond dimension enforces a finite correlation length. This in turn means that we cannot provide a precise estimate of the exponent $\nu_\perp$. As a reference value, we can fit our results as done for obtaining $\beta$ --- in this case neglecting the points showing a rounding of the divergence --- which yields $\nu_\perp= 0.5\pm0.2$.

\noindent {\bf \em Summary and conclusions -}
We have found strong evidence that the QCP in $1d$ features a continuous absorbing-state phase transition. Estimates for the critical exponents are summarized in Table~\ref{FigR6}. While clearly different from those of DP universality, they are remarkably close to those predicted for the tri-critical point of a mixed quantum and classical branching process in $2d$ \cite{Buchhold2017}. This might be a coincidence but could be rationalized in the following way: Ref.~\cite{Buchhold2017} predicts the emergence of tri-critical point for a contact process in which quantum and classical branching compete, and a first-order transition when only quantum branching is present. As, according to our findings, there is apparently no first-order transition in $1d$ this could mean that quantum fluctuations (which are to a large extent neglected in \cite{Buchhold2017}) shift the tri-critical point onto the quantum axis, i.e. its occurrence does not require additional classical processes. However, the mismatch in the dimension, i.e. $1d$ vs. $2d$,  remains puzzling: it would be interesting to understand whether this could be due to an underlying mapping of such a $1d$ quantum system to a $2d$ classical one, as it can occur to equilibrium systems.
\begin{table}
  \centering
  \begin{tabular}{|c|c|c|c|c|}
  \hline
    &QCP & $1d$ DP \cite{Hinrichsen2000b}& $2d$ DP \cite{Hinrichsen2000b}& $2d$ Ref.~\cite{Buchhold2017} \\
    \hline
    $\delta$ &0.36&0.16&0.45&0.35\\
    \hline
    $\beta$&0.36&0.28&0.58&0.35\\
    \hline
    $\nu_\parallel$&1.00&1.73&1.30&1.03\\
    \hline
    $\nu_\perp$&0.5&1.10&0.73&0.52\\
    \hline
  \end{tabular}
  \caption{Exponents of the QCP as well as of the $1d$ and $2d$ DP \cite{Hinrichsen2000b}. The last column displays exponents for the tri-critical point of a quantum plus classical branching $2d$ process \cite{Buchhold2017}.} \label{FigR6}
\end{table}


\begin{acknowledgments}
\noindent {\bf \em Acknowledgements -} We thank Maryam Roghani and Matteo Marcuzzi for 
fruitful discussions. The research leading to these results has received funding from the European Research Council under the European Unions Seventh Framework Programme (FP/2007-2013)/ERC [grant agreement number 335266 (ESCQUMA)], the Engineering and Physical Sciences Council [grant numbers EP/M014266/1 and EP/R04340X/1], the Leverhulme Trust [grant number RPG-2018-181], the 
Volkswagen
Foundation and the DFG within EXC 2123 (QuantumFrontiers), SFB 1227
(DQ-mat), and SPP 1929 (GiRyd). IL gratefully acknowledges funding through the Royal Society Wolfson Research Merit Award. 
\end{acknowledgments}
\bibliographystyle{apsrev4-1}
\bibliography{QCPRef}


\end{document}